\documentclass[%
reprint,
amsmath,amssymb,
aps,
]{revtex4-2}

\usepackage{graphicx}% Include figure files
\usepackage{dcolumn}% Align table columns on decimal point
\usepackage{bm}% bold math
\usepackage{xcolor}
\begin{document}
	
	\preprint{APS/123-QED}
	
	\title{Dielectric functions of $\alpha$Sn: insight from the first principles}% Force line breaks with \\
	%\thanks{A footnote to the article title}%
	
	\author{J. S. Duan}
	\affiliation{
		KBR, Inc., Beavercreek, OH 45431, USA \\ Department of Physics, University of Dayton, OH 45469}
	%	 \altaffiliation[Also at ]{Department of Physics, University of Dayton, OH 45469, USA}%Lines break automatically or can be forced with \\
	\email{j.samuel.duan@gmail.com}
	
	\date{\today}% It is always \today, today,
	%  but any date may be explicitly specified
	
	\begin{abstract}
		Gray tin ($\alpha-$Sn) is a group-IV, zero-gap semiconductor with potential applications in electronic devices, necessitating a clear understanding of its dielectric properties.  We report the first-principles calculations of the band structure and dielectric function of both unstrained and strained $\alpha$-Sn using the density functional theory, emphasizing an interesting absorption at $\sim$ 0.41 eV measured in the spectroscopic ellipsometric experiment. \cite{carrasco2018direct} We found that this optical absorption arises from the interband transition $\Gamma_{8v}^{+} \rightarrow \Gamma_{7c}^{-}$. This result could provide a foundation for further study of the spin-polarized photocurrent in strained $\alpha-$Sn, a topological elemental material, and for its applications for spintronics.

		%\begin{description}
		%\item[Usage]
		%Secondary publications and information retrieval purposes.
		%\item[Structure]
		%You may use the \texttt{description} environment to structure your abstract;
		%use the optional argument of the \verb+\item+ command to give the category of each item. 
		%\end{description}
	\end{abstract}
	
	%\keywords{Suggested keywords}%Use showkeys class option if keyword
	%display desired
	\maketitle
	
	%\tableofcontents
	
	\section{\label{sec:level1}INTRODUCTION}

Major goals in current electronic devices are to increase the speed of electronic devices from the gigahertz to the terahertz range and to extend their functionalities by electron spin. \cite{braun2016ultrafast} In these respects, topological insulators are promising materials for revolutionizing spintronics and other fields of emerging electronics.\cite{gladczuk2021study} The most remarkable properties of their helical surface states is spin-velocity locking that protects forbidden backscattering, ensures full spin polarization, the exceptional transport mobilities, and low energy dissipation, that is very attractive for semiconductor devices in high-speed communication applications. \cite{braun2016ultrafast} The prerequisites of these properties is large spin polarization induced by means of surface currents. Illuminated by circularly polarized light, spin-polarized surface currents are excited via interband transitions between the occupied and unoccupied helical topological surface states restricted by angular momentum selection rule.\cite{mciver2012control, bai2022ultrafast}
	$\alpha-$Sn (gray tin) is the diamond-crystalline phase of elemental tin and a so-called  ``zero-gap semiconductor''.\cite{ewald1958gray}  Owing to its unusual band structure, extensive investigations were carried out primarily in the 1960s on bulk samples \cite{groves1963band,cardona1966electroreflectance, pollak1970energy} and in the 1980s and 1990s on epitaxial films. \cite{tu1989growth,john1989core}  More recently, the prediction and verification of $\alpha-$Sn as a topological insulator (under strain) \cite{barfuss2013elemental}, has advantages over the established and experimentally confirmed binary ($\rm e.g. Bi_{x}Se_{y}$) or ternary topological materials ($\rm e.g. Pb_{1-x-y}Bi_{x}Se_{y}$) that have the intrinsic problem of defect-induced bulk conductivity. Elemental tin, in contrast, is a much simpler system to be fabricated without a well-controlled composition requirement. It can be integrated into established compound semiconductor devices due to the substrate compatibility and quasi-lattice-matched to, for example, InSb (6.4798 Å).
	
	Of primary interest for device applications is the recent observation of an optical transition in the infrared region at $\sim$0.41 eV ($\sim 2.7\mu m$) in $\alpha-$Sn using spectroscopic ellipsometry that corresponds with a `negative' band gap associated with an optical transition $\Gamma_{8v}^{+} \rightarrow \Gamma_{7c}^{-} $.\cite{carrasco2018direct} This interband optical transition could be asymmetric optical excitation from the occupied to the unoccupied states across the Dirac point in the helical Dirac cone, producing spin-polarized surface current. Therefore, understanding this transition has an important meaning for unprecedented applications. 
	However the exact origin of this transition is not fully understood yet, calling for a renewed effort for a first-principles calculation of the optical properties which can be determined largely from its electronic band structure.	
	
	The band structure of $\alpha-$ Sn has many unusual features, including a degeneracy at the $\Gamma$-point that closes the gap between the valence and conduction bands.  Because of the strong relativistic Mass-Darwin effect on the $\alpha$-Sn in 5s electrons, the $\Gamma_{7c}$ level moves down in the energy and dips below the $\Gamma_{8v}$ level, resulting in band inversion. \cite{kufner2013structural} Moreover, the strain and spin-orbit interaction influence the band structure considerably.\cite{reno1990stability} The band gap at the $\Gamma$-point can be opened, however, through epitaxial strain or by alloying with Ge. In-plane compressive strain ($-0.164\%$) in $\alpha-$Sn grown on InSb substrate can lift the band degeneracy at the $\Gamma$ point creating a small band gap of 30 meV.\cite{barfuss2013elemental}  Careful consideration of many factors are needed to create an accurate fundamental model that manifests these several unusual features.
	
	In this paper, we report the first-principles calculation of the band structure and optical properties of $\alpha-$ Sn. The emphasis is on the optical transition in the infrared region and the effect of the spin-orbit interaction and strain on the band structure and optical properties. Our  calculations provide insight into the fundamental physics in the optical properties of $\alpha$-Sn in the infrared region.
	
	The paper is organized as follows. In section \ref{sec:method}, we describe the theory and computational methods. In section \ref{sec:results}, we present the calculated band structure and optical properties. The results are compared with previous calculations.
	
	\section{\label{sec:method} Methods}
	
	Density functional theory (DFT) has been applied as the standard method to investigate the properties of solids since it was initially proposed more than 50 years ago. It has been justified and validated that well-tuned pseudo-potentials can reproduce very precisely all-electron results for many properties of solids, but at a much lower computational time. By calculating the band gaps of a large data set of 473 solids with norm-conserving pseudopotentials and comparing with all-electron calculations, Borlido {\it et. al.} found the results are encouraging that on average, the absolute error is about 0.1 eV, yielding absolute relative errors in the 5-10 \% range. \cite{borlido2020validation} DFT and pseudo-potentials have been applied increasingly to study more complex system with techniques requiring massive computational efforts such as the many-body perturbation theory, spin-orbit interaction, and materials genome. \cite{kucukbenli2014projector}

	To guarantee the accurate computational results using the exchange-correlation functionals, careful consideration and tests are essential to ensure the quality for simulating the properties of topological insulators.
	
	To address the electronic and optical properties of topological insulators, two important factors need to be considered. The first is spin-orbit interaction. Topological surface states arise from spin-orbit interaction that is strong enough to change the band topology. \cite{che2020strongly} The second is many-body problem. In the DFT formalism, the ground state properties of a many-electron system are determined from the electron density distribution alone, which is described by a single determinant with all many-body effects included in one term, the exchange-correlation functional, avoiding massive computation of complex many-dimensional wave functions. \cite{Flores2018accuracy} It is very important to validate and justify the appropriate the exchange-correlation functional describing the exchange correlation in many-electron systems.
	
	All first-principles calculations are  performed with the Vienna $\it ab-initio$ simulation package (VASP) (Version 5.4).\cite{kresse2001vienna} As mentioned above, the first-principles simulation results are sensitive to the exchang-correlation functionals chosen for a specific problem. To this end, the exchange-correlation is modeled with local density approximation (LDA) and hybrid functional (HSE). \cite{blochl1994projector} We set the energy cutoff of the plane wave basis set to 500 eV. The frequency dependence of the dielectric function is calculated using the independent particle approximation provided within the VASP package. The $32 \times 32 \times 32$ $\mathbf k$-points are used in LDA calculation, while $16 \times 16 \times 16$ $\mathbf k$-points are used in HSE calculation using an averaging over multiple grids method. \cite{bokdam2016role} 
	
	To incorporate strain into the band structure calculation, the deformation of the unit cell is set using the theory of linear elasticity. We explicitly calculate the experimentally relevant case of $\alpha-$Sn grown epitaxially on an InSb (001) substrate. The lattice parameter of $\alpha-$Sn and InSb is 6.4892 $\AA$ (20 $^{\circ}$C ) \cite{oehme2011room} and 6.4794 $\AA$ (25 {$^{\circ}$}C),\cite{straumanis1965lattice} respectively. Due to the lattice mismatch, the InSb sublattice introduce in-plane compressive strain to the $\alpha-$Sn film on the top of it. First-principles calculation is performed at 0 K. To simulate the strain effect, the volume of crystal is fixed by changing the ratio of lattice constant in the x-y axis, $a_{\parallel}$, and that in z axis, $a_{\perp}$. The $\alpha$ phase of Sn has the cubic diamond crystal structure with a lattice parameter of  $a=6.4746 \AA$ at 0 K estimated with LDA and $a=6.5801 \AA$ estimated with HSE functional. Indeed, the band gap and optical property of a material can be sensitive to the lattice parameters. The electronic and optical properties of $\alpha$-Sn are calculated in such a way that the geometry optimization is performed with LDA approximation, which predicts the lattice parameter of $\alpha$-Sn, 6.4794 $\AA$, with a reasonable error of -0.225\% with respect to the experimental value, followed by a single point calculation of electronic and optical properties using using either the LDA or HSE funcationals.
	
	When a film of $\alpha-$Sn is coherently constrained to the InSb substrate lattice, it has a biaxial in-plane compressive strain, $\epsilon_{\parallel}$ of $-0.164\%$, which is chosen to represent a reasonable strain effect and does not reflect the real value of the compressive strain because the linear thermal expansion fails when the temperature approaches 0 K. Thus, the corresponding lattice constant in the x-y axis $a_{\parallel}=6.4640\AA$. Assuming a linear elastic response, the out-of-plane strain is then $\epsilon_{\perp}=-2(C_{12}/C_{11})\cdot\epsilon_{\parallel}=0.00138$, where $C_{11} = 0.69$ GPa and $C_{12}=0.29$ GPa, \cite{moontragoon2007band} resulting in an out-of-plane lattice constant of $a_{\perp}=6.4836 \AA$. 
	
	It is worth mentioning that the deformation due to a bi-axially symmetric, in-plane strain is different from a uni-axial strain applied along the orthogonal direction.  In the former case, $\epsilon_{\perp} = (-2C_{12}/C_{11}) \epsilon_{\parallel}$; in the latter case,  $\epsilon_{\perp} = -[(C_{12}+C_{11}/C_{11}] \epsilon_{\parallel}$.  Our strain-dependent results should not be confused with calculations by others using different boundary conditions and definition of the compressive and tensile strain.\cite{hinckley1990influence,ruan2016symmetry}  
	
	\section{\label{sec:results}Results}
	
	We discuss the first-principles calculation of the band structure $\alpha-$Sn in section \ref{sec:band} and its dielectric function in section \ref{sec:dielectric} focusing on the infrared transition and the influences of the spin-orbit interaction and in-plane compressive strain. 
	
	\subsection{\label{sec:band} Electronic band structure of $\alpha$-Sn}
	
	The first-principles calculation are performed on $\alpha$-Sn to understand the relative influences of spin-orbit interaction and strain on electronic band structure of $\alpha-$Sn regarding relative energy of bands. Of particular interest is the $\bar{E}_{0}$ transition at $\sim$ 0.41 eV in our ellipsometric measurements. Previous experimental measurements typically cut-off above 1 eV \cite{vina1984temperature}; Kufner $et$ $al$. calculated the dielectric function of $\alpha-$Sn using HSE hybrid functional, and compared the results with and without the spin-orbit interaction. The $\bar{E}_{0}$ peak is not pronounced in the calculated dielectric function. \cite{kufner2013optical}  Pollak $et$ $al$. calculated the band structure of $\alpha-$Sn and assigned $\bar {E}_{0}$ of 0.41 eV to the transition between $\Gamma_{8}^{+}  \rightarrow  \Gamma_{7}^{-}$. This transition has not been considered in any great detail. 
	
	\begin{figure}[h]
		\centering
		\includegraphics[width=8.6cm, height=5.0cm]{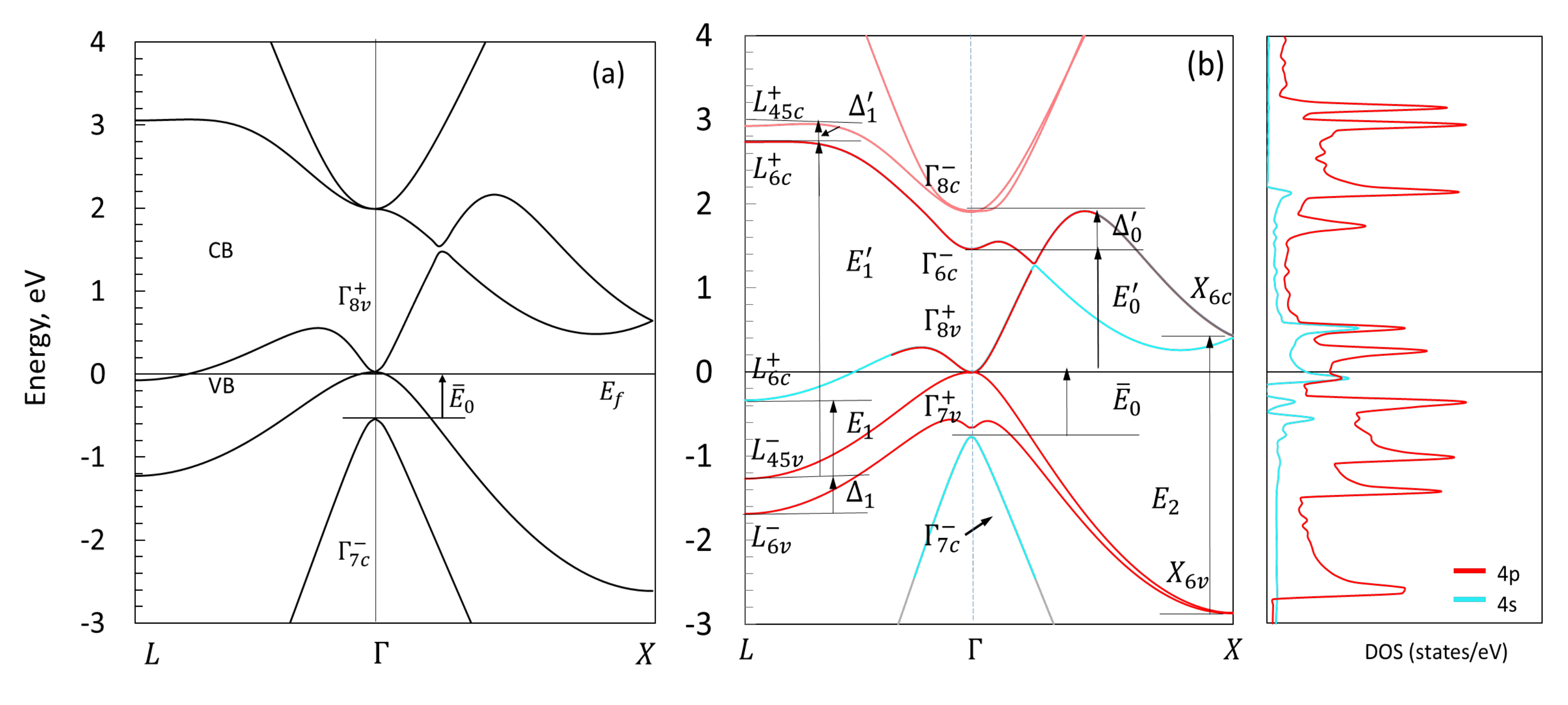}
		\caption{Band structure of $\alpha-$Sn calculated with LDA, (a) without, and (b) with  spin-orbit interaction. The orbital character is encoded by a cyan and red color code, respectively, for 4$s$ and 4$p$  orbital character in (b).
		}
		\label{fig:snsoi-PBE-01-15-20-b}
	\end{figure}
	
	To understand the origin of the $\bar{E}_{0}$ transition, we calculate the band structure with and without the spin-orbit interaction, and the density of states of $\alpha-$Sn around the $\Gamma$ point, and present the results in Figure \ref{fig:snsoi-PBE-01-15-20-b}. The bands are labeled using the notation of Pollak {\it et al.} \cite{pollak1970energy} and  $+$, $-$ superscripts denote the parity. Optical transitions are determined using selection rules, meaning transitions can only connect states with opposite parities.\cite{peter2010fundamentals} 
	
	Without spin-orbit interaction, the transition $\bar{E}_{0}$ occurs between $\Gamma_{8v}^{+}$ and $\Gamma_{7c}^{-}$, which is lower than the valence band $\Gamma_{8v}^{+}$, demonstrating an inversed band structure shown in Figure \ref{fig:snsoi-PBE-01-15-20-b} (a). The band structure (the position of light hole and heavy hole) of $\alpha-$Sn is modified by spin-orbit interaction, shown in Figure \ref{fig:snsoi-PBE-01-15-20-b} (b). $\Gamma_{7v}^{+}$ subband is split from the valence band $\Gamma_{8v}^{+}$ and $\Gamma_{7c}^{-}$ band is shifted down by spin-orbit interaction.\cite{barfuss2013elemental} This is caused by the quantum mechanical repulsion that the $\Gamma^{+}_{8v}$ and $\Gamma_{7c}^{-}$ band are pushed apart by the split-off band $\Gamma_{7v}^{+}$.\cite{chao1992spin}
	
	\begin{figure}
		\centering
		\includegraphics[width=8.2cm, height=5.86cm]{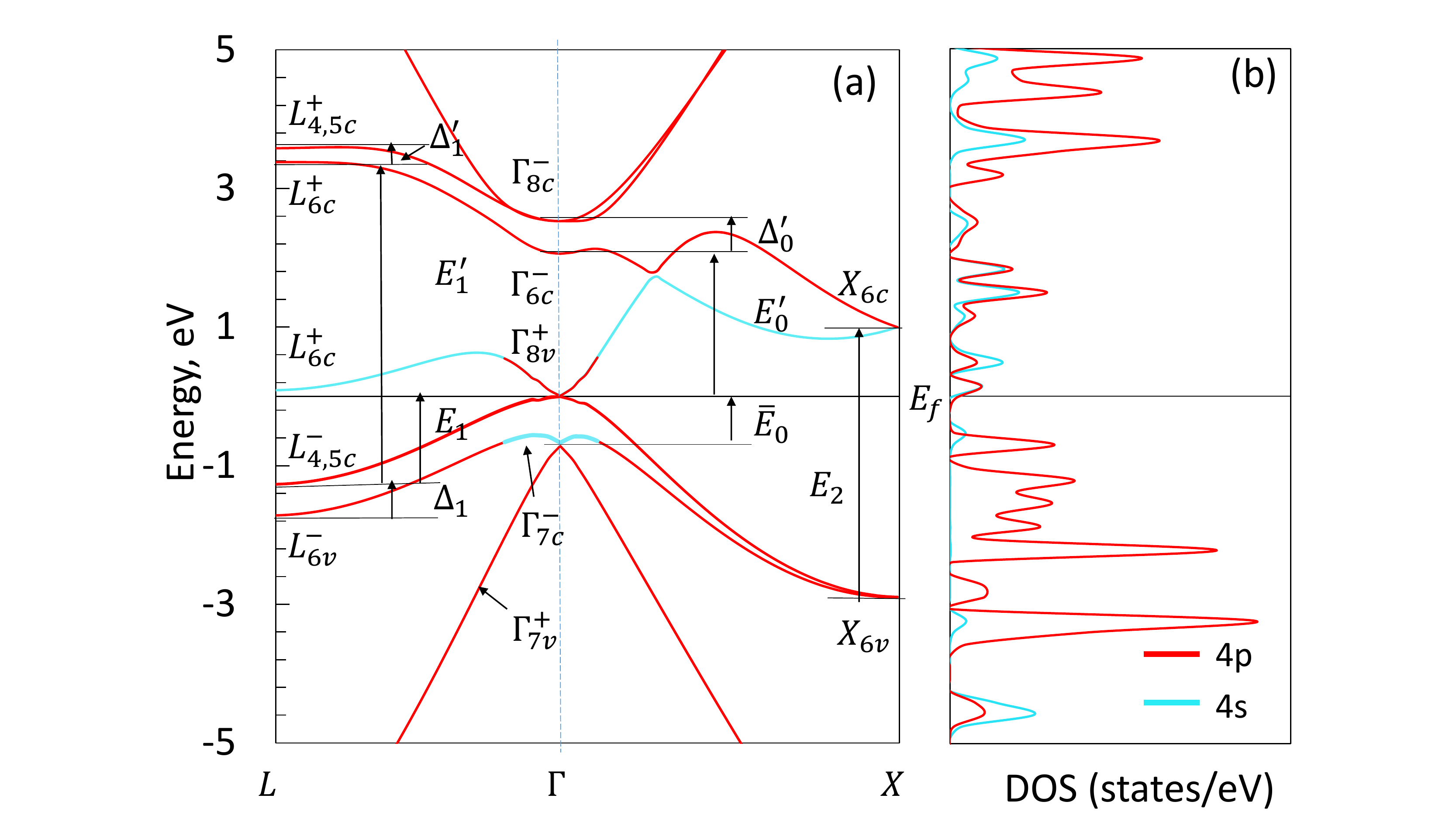}
		\caption{(a) Band structure of $\alpha$-Sn and (b) projected density of states calculated with spin-orbit interaction and HSE 06 hybrid functional. The orbital character is encoded by a cyan and red color code, respectively, for 4$s$ and 4$p$  orbital character in (a) and (b).
		}
		\label{fig:snsoi-HSE-band-05-15-19}
	\end{figure}
	
	\begin{figure}
		\centering
		\includegraphics[width=8.2cm, height=5.86cm]{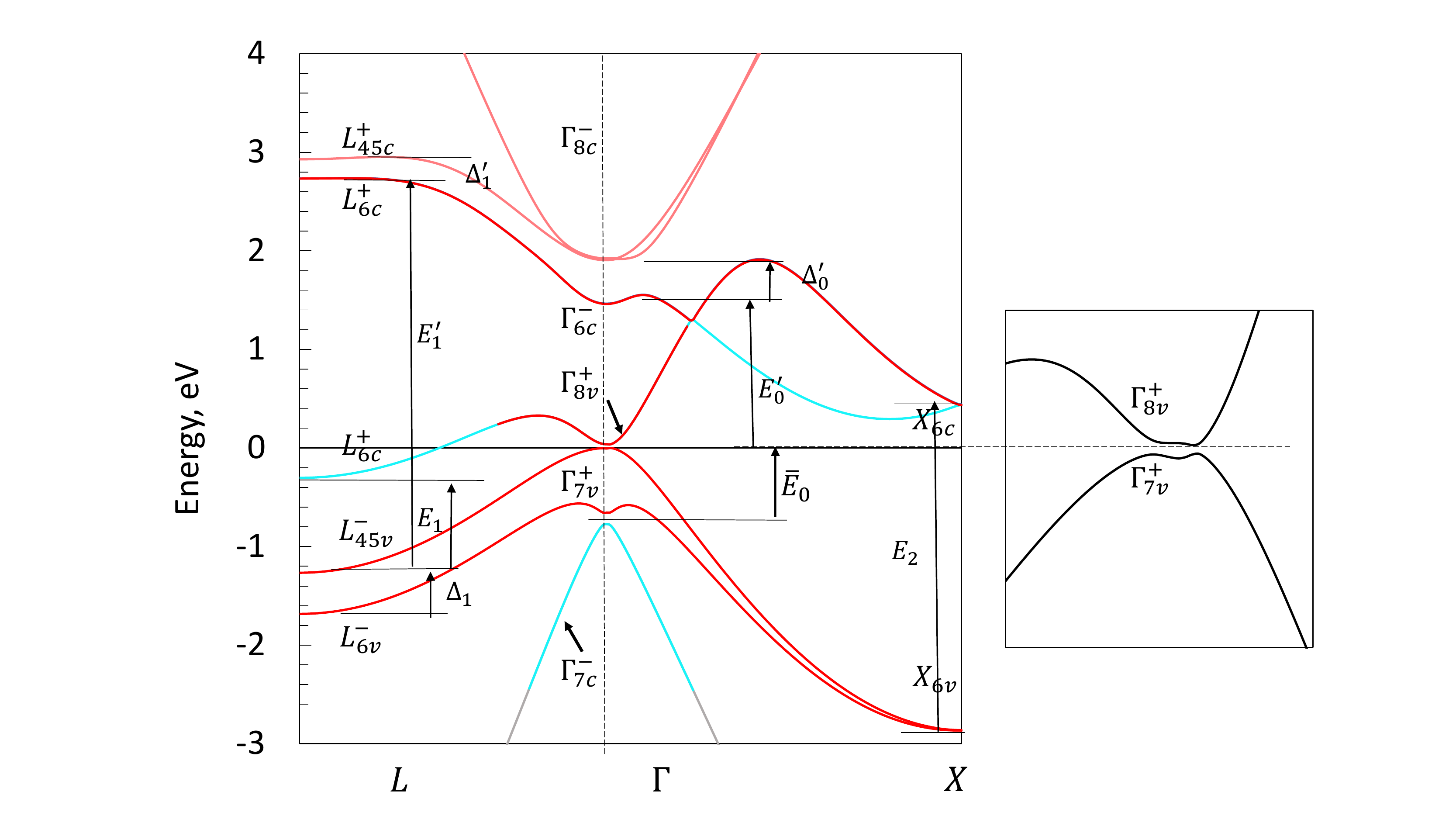}
		\caption{Band structure of $\alpha$-Sn deformed with $-0.164\%$ in-plane compressive strain calculated with LDA.
		}
		\label{fig:sn-snsoi-PBE-band-strain-07-23-22}
	\end{figure}
	
	\begin{table}
		\caption{Band gap (in eV) in $\alpha-$Sn calculated using LDA and HSE hybrid functional with spin-orbit interaction (SO) compared with computational data reported in literature. }	
		\label{table:band} 
		
		\begin{center}
			\begin{ruledtabular}
				\begin{tabular}{ c c c c c c  c }	
					Energies     & Transitions          & \multicolumn{3}{c}{This work w.SO}   &  $\mathbf{k \cdot p}$ \footnotemark[1] & HSE \footnotemark[2] \\ \cline{3-5}
					
					&    & LDA    & LDA  \footnotemark[3]   & HSE     &          &                        \\                
					\hline 
					$\bar{E}_{0}$     &  $\Gamma_{7c}^{-} \rightarrow \Gamma_{8v}^{+}$       &  0.69  & 0.70  & 0.58    & 0.40   & 0.52 \\
					$E_{1}$           &  $\Gamma_{45v}^{-} \rightarrow \Gamma_{6c}^{+}$           &  0.84  & 0.80  & 1.00    & 1.10   &  $\mathrm{-}$           \\      
					$E_{1}+\Delta_{1}$    &  $\Gamma_{6v}^{-} \rightarrow \Gamma_{6c}^{+}$       &  1.28   & 1.24  & 1.48    & 1.55   &  $\mathrm{-}$           \\
					$E_{2}$        &  $\Gamma_{6v}^{-} \rightarrow \Gamma_{6c}^{+}$              &  3.24   & 3.11  & 3.73    & 3.25   &  $\mathrm{-}$            \\
				\end{tabular}
			\end{ruledtabular}
			\footnotetext[1]{The calculated data from Ref.~\cite{pollak1970energy}}
			\footnotetext[2]{The calculated data from Ref.~\cite{kufner2013structural}}
			\footnote[3]{The data calculated with in-plain compressive strain of $-0.164\%$}
		\end{center}
	\end{table}

	$\alpha-$ Sn demonstrates the inverted band structure that the conduction band $\Gamma_{7c}^{-}$ at the $\Gamma$ point is lower than the valence band $\Gamma_{8v}^{+}$, as shown in Figure \ref{fig:snsoi-PBE-01-15-20-b} and \ref{fig:snsoi-HSE-band-05-15-19}. \cite{groves1963band} The possible transitions are indicated in these Figures and listed in Table \ref{table:band}. In Figure \ref{fig:snsoi-PBE-01-15-20-b}(b), the band ordering calculated with LDA is $\Gamma_{8v}^{+} > \Gamma_{7v}^{+} > \Gamma_{7c}^{-}$, where $\Gamma_{7v}^{+}$ is $4p$ orbital character, and is a split-off valence band from $\Gamma_{8c}^{+}$ which is mixed $4s$ and $4p$ orbital character, but is $4p$ orbital character near the $\Gamma$ point shown in red line. $\Gamma_{7c}^{-}$ is a conduction band of $4s$ orbital character, which is lower than the valence band $\Gamma_{8v}^{+}$ by 0.69 eV corresponding to $\bar{E}_{0}$ interband transition.
	
	The $\bar{E}_{0}$ band gap calculated using LDA with and without the spin-orbit interaction is 0.70 eV and 0.28 eV, respectively. Without the spin-orbit interaction, the band gap $\bar{E}_{0}$ estimated using the LDA is predicted correctly to be negative ($\Gamma_{7c}^{-}$ is lower than ($\Gamma_{8v}^{+}$), but is much lower than the experimental value 0.41 eV. That can be that the LDA underestimates the band gap due to the inaccurate description of the exchange correlation.\cite{giovannetti2015kekule} With the spin-orbit interaction, the $\Gamma_{7c}^{-}$ is pushed down far to 0.70 eV, leading to an overestimated band inversion. LDA produces higher spin-orbit splitting,\cite{scherpelz2016implementation} it could be because a purely perturbative treatment of spin-orbit coupling is not sufficient in the LDA approximation. \cite{sakuma2011g}
	
	HSE has been applied to calculate the band gaps of covalent materials including semiconductors, \cite{janesko2009screened} and has much more computational efficient than other many-body perturbation methods, and provides performances well in agreement with G0W0 or GW calculations.\cite{giovannetti2015kekule}	With the spin-orbit interaction, HSE improved the band gap $\bar{E}_0$ to 0.58 eV, better than LDA does, likely because incorporation of a fraction of the exchange within the HSE functional reduces self-interaction errors, leading to a better result. \cite{ramasubramaniam2012large}
	The spin-orbit splitting is sensitive to the level of theory employed. As a result, the band order can be dependent on the exchange-correlation functional applied in the calculations. \cite{ramasubramaniam2012large}
	The band order is $\Gamma_{8v}^{+} > \Gamma_{7c}^{-} > \Gamma_{7v}^{+}$ calculated with HSE hybrid functional as shown in Figure \ref{fig:snsoi-HSE-band-05-15-19}, which is consistent with $\mathbf{k \cdot p}$ calculation, \cite{pollak1970energy} but is different from the results obtained with LDA, where split-off band $\Gamma_{7v}^{+}$ is higher than $\Gamma_{7c}^{-}$. $\bar{E}_{0}$ transition occurs between $\Gamma_{7c}^{-}$ and $\Gamma_{8v}^{+}$. Our value is 0.58 eV which is close to 0.52 eV reported in \cite{kufner2013structural}. Both calculations using LDA and HSE hybrid functional show that the $\bar{E}_{0}$ is interband transition between inverted the valence band $\Gamma_{8_{v}}^{+}$ and $\Gamma_{7c}^{-}$ conduction band.
	
	The calculated band structure of $\alpha-$Sn under $-0.164$\% biaxial compressive strain is presented in Figure \ref{fig:sn-snsoi-PBE-band-strain-07-23-22} . The symmetry from the $O_{h}$ (cubic) unit cell to $D_{4h}$ (tetragonal) caused by strain, see Figure S12 the supplementary material in \cite{carrasco2018direct}. The band structure in Figure \ref{fig:sn-snsoi-PBE-band-strain-07-23-22} refers to the direction \textit{perpendicular} to the plane of applied strain,\textit{i.e.} the surface plane of the film.  
	
	The in-plane compressive strain changes the band structures. The most notable feature is that the strain opens a gap between the conduction band $\Gamma_{8c}^{+}$ and valence band $\Gamma_{8v}^{+}$ at the $\Gamma$ point, \cite{yang2004si} as shown in the inlay in Figure \ref{fig:sn-snsoi-PBE-band-strain-07-23-22}. The magnitude of the gap opening depends on the computation method. The LDA functional gives 31 meV, which is closer to the value of 30 meV calculated under a in-plane compressive strain of $-0.14$ \% using QW quasi-particle calculation, \cite{barfuss2013elemental} but larger than 13 meV calculated under the in-plane strain of $-0.13\%$ using Pikus-Bir Hamiltonian.\cite{carrasco2018direct} 
	
	To gain insight into how the band structure changes with strain, we consider the general Luttiger Hamiltonian of an electron (or hole) in a semiconductor including only the four low-energy bands around the $\Gamma-$point.\cite{luttinger1956quantum, zhang2018engineering}
	
	The four low-energy bands of strained $\alpha-$Sn around the $\Gamma$ point can be written as, \cite{ruan2016symmetry}
	
	\begin{equation}
		H= H_{L} + H_{strain},
	\end{equation}
	
	where $H_{strain}$ is an additional perturbation term $H_{strain}$ created by externally applied strain and written as follows,
	
	\begin{equation}
		H_{strain}=-g \Big( J_{z}^{2}-\frac{5}{4}\Big),
		\label{h-strain}
	\end{equation}

	where $g$ is determined by the strength of the strain and $g > 0 $ for biaxial compressive strain. 
	
	$\alpha-$Sn is an inverted band structure materials, where the light hole band $\Gamma_{8v}^{+}$ moves up become unoccupied. Under the biaxial compressive strain, the light hole ($J_{z}=\pm \frac{1}{2}$) band, $\Gamma_{8v}^{+}$, is pushed up ($H_{strain} > 0$, and $H$ increases), while the heavy hole band, $\Gamma_{8c}^{+}$ ($J_{z}=\pm \frac{3}{2}$), is pushed downward ($H_{strain} < 0$, and $H$ decreases). A small gap thus opens between $\Gamma_{8c}^{+}$ and $\Gamma_{8v}^{+}$ under a in-plane compressive strain. This result quantitatively agrees with the computational and theoretical prediction, as shown in Figure \ref{fig:sn-snsoi-PBE-band-strain-07-23-22}. 
	
	The magnitudes of optical transitions such as $\bar{E}_{0}$, $E_{1}$, and $E_{2}$ also are changed between 8.5 meV and 25 meV in $\alpha-$Sn under $-0.164\%$ in-plane compressive strain. This result is in consistent with the energy peaks estimated from the calculated dielectric function and our ellipsometry experiment discussed in \ref{sec:dielectric}. 
	
	The strained $\alpha$-Sn is a topological elemental material, where the topological surface states are formed between $\Gamma_{7c}^{-}$ and $\Gamma_{8v}^{+}$, as illustrated in Figure \ref{fig:asn-surface-08-01-22} in the appendix \ref{sec:appendix-0}.  The direct interband transition $\bar{E}_{0}$  in the topological surface states, from $\Gamma_{8v}^{+}$ to the  $\Gamma_{7c}^{-}$ across the Dirac point, resulting in spin polarized photocurrents, which are very important for developing $\alpha$-Sn based spintronics.
	
	\subsection{\label{sec:dielectric} Dielectric function of $\alpha-$Sn}
	
	Once the band structure of $\alpha$-Sn is calculated, the frequency dependent dielectric function due to interband transitions can be calculation as follows.
	The imaginary part of the frequency dependent dielectric constant of $\alpha$-Sn, $\epsilon''(\omega)$, is calculated within the independent-particle approximation and is expressed as, \cite{adolph1996nonlocality, grosso2000optical, gajdovs2006linear}
	
	\begin{eqnarray}
		\epsilon''_{\alpha \beta }(\omega)=&& \frac{4\pi^{2}e^{2}}{\Omega} \lim_{q\rightarrow 0} \frac{1}{q^{2}} \sum_{c,v,\mathbf{k}}2w_{\mathbf{k}}
		\delta(\epsilon_{c\mathbf{k}}-\epsilon_{v\mathbf{k}}-\omega) \nonumber \\
		&& \times <u_{c\mathbf{k+e}_{a}q}|u_{v_\mathbf{k}}>
		<u_{c\mathbf{k+e}_{\beta}q}|u_{v\mathbf{k}}>^{\ast},
		\label{eq:dielectric}
	\end{eqnarray}
	
	where $\Omega$ is the volume of the primitive cell, $\mathbf{q}$ the Bloch vector of the incident wave, $\mathbf{k}$ a wavevector within the first Brillouin zone, the indices c and v the conduction and the valence band states, respectively, $w_{\mathbf{k}}$ the $k$ point weights, the unit vectors $\mathbf{e}_{\alpha}$ for the three Cartesian directions.  
	
	Subsequent calculation of the real part of the dielectric function, $\epsilon'(\omega)$, is done by applying the Kramers-Kronig relationship.\cite{gajdovs2006linear} 
	
	\begin{eqnarray}
		\epsilon_{\alpha \beta}'(\omega)=1+\frac{2}{\pi}P\int_{0}^{\infty}\frac{\epsilon_{\alpha \beta}''(\omega ')\omega'}{\omega'^{2}-\omega^{2}} d \omega',
	\end{eqnarray}
	
	where $P$ denotes the principal value.
	
	\begin{figure}[h]
		\centering
		\includegraphics[width=8.2cm, height=5.88cm]{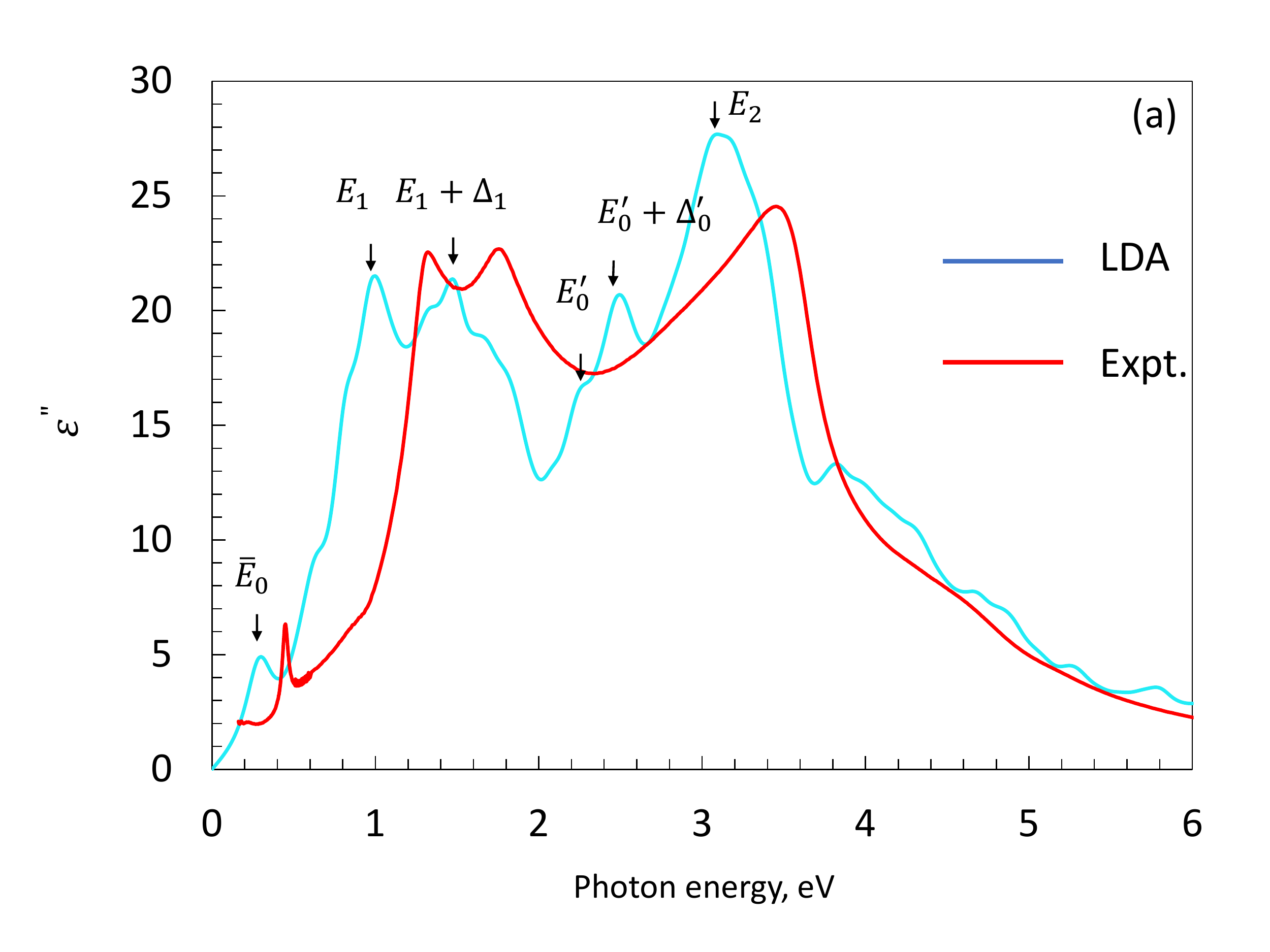}
		\includegraphics[width=8.2cm, height=6.2cm]{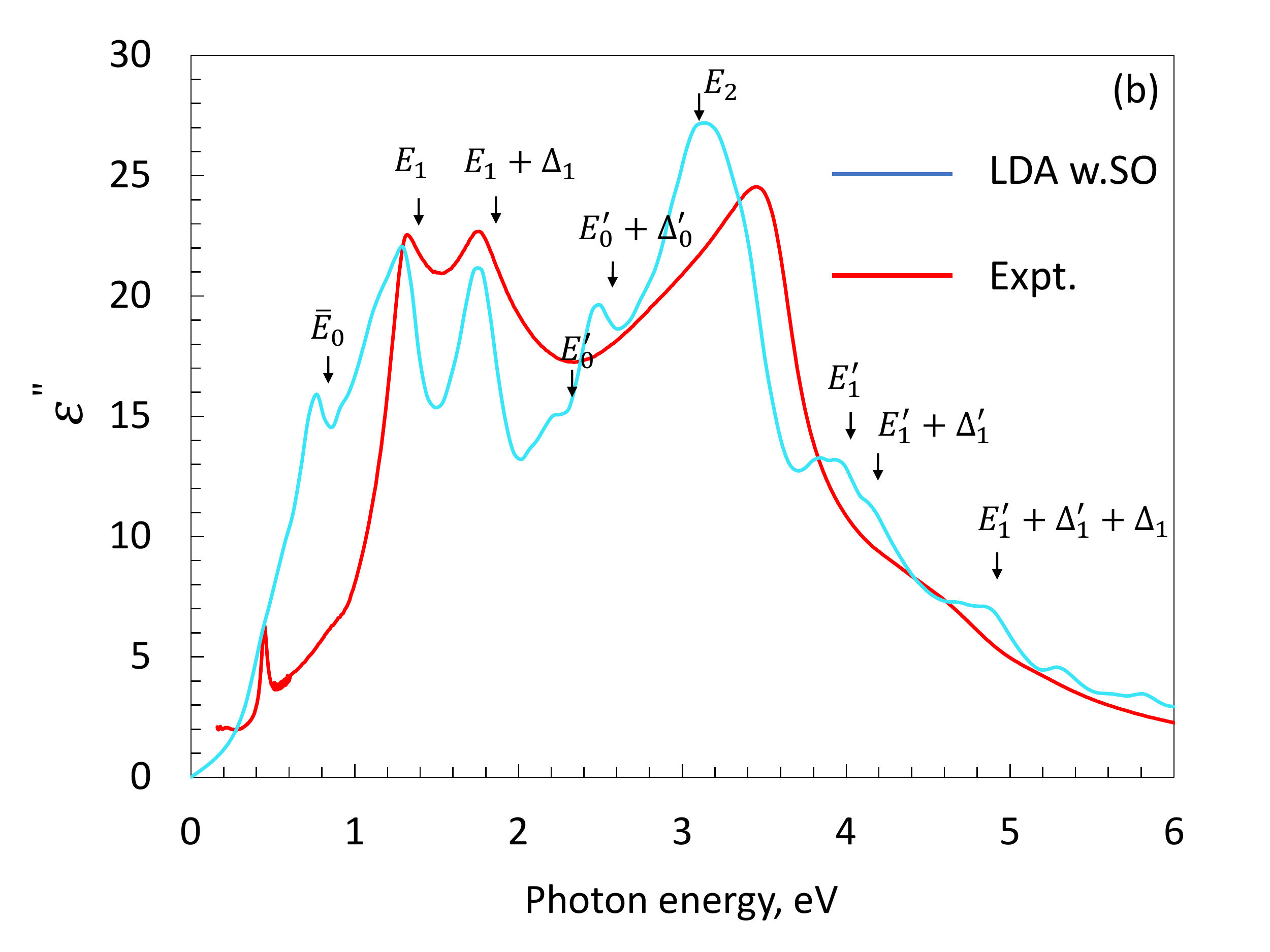}
		\caption{Computed imaginary part of the dielectric function of $\alpha$-Sn using LDA functional without (a) and with (b)  the spin-orbit interaction shown in cyan together with the experimental data shown in red. 
		}
		\label{fig:asnldaso-6-22-18-w-exp}
	\end{figure}

	\begin{figure}[h]
		\centering
		\includegraphics[width=8cm, height=6cm]{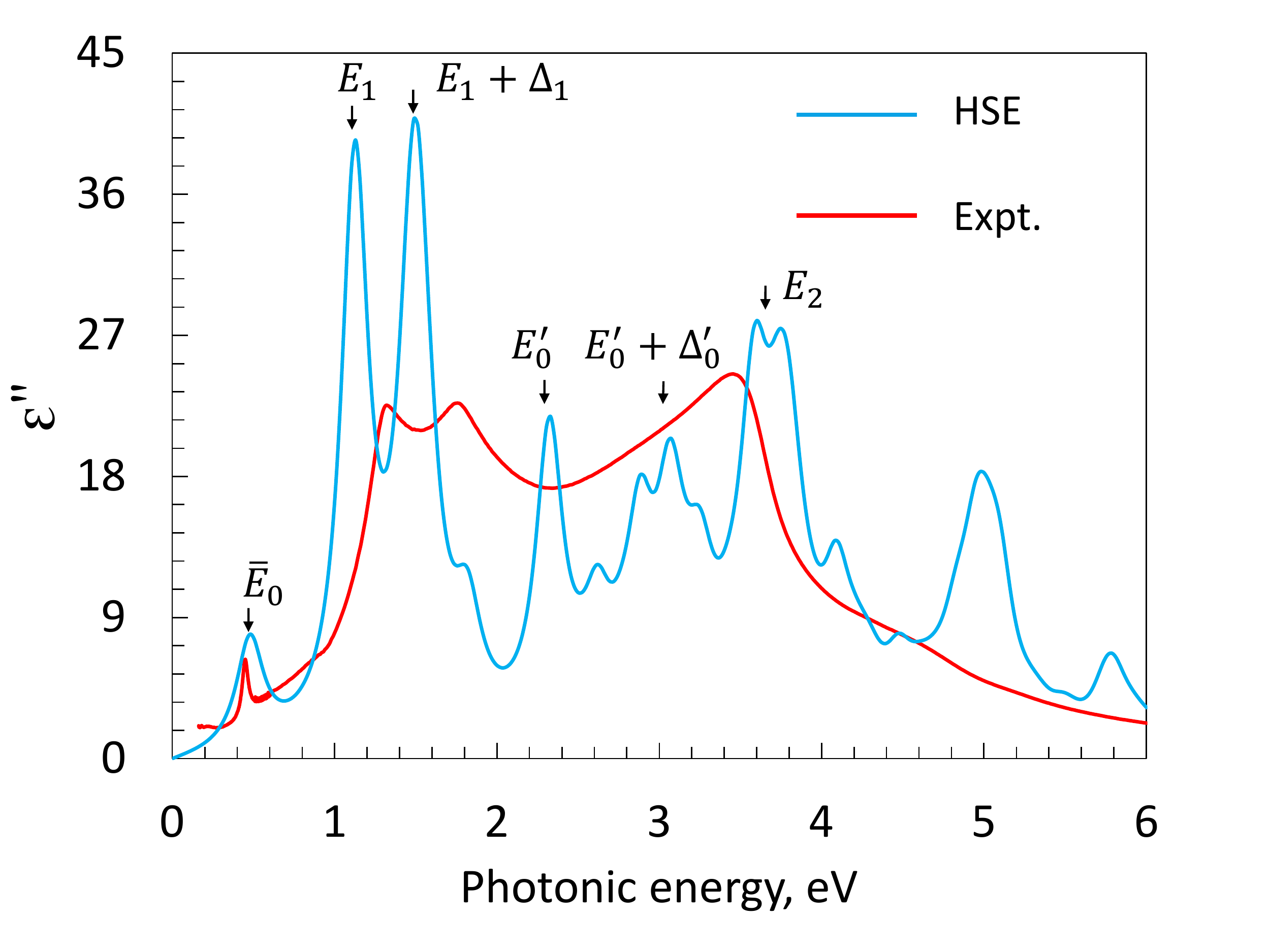}
		\caption{Computed imaginary part of dielectric function of $\alpha$-Sn using HSE hybrid functional shown in the cyan line compared with the experimental data shown in the red line. \cite{carrasco2018direct} 
		}
		\label{fig:asn-hse-exp-190528}
	\end{figure}
	
	Since the imaginary part of the dielectric function is calculated based on the interband transitions, there is a good correlation between the energy peaks in the dielectric function and the interband transitions, making the assignment of energy peaks to the interband band transitions straightforward. \cite{albanesi2000electronic} The main optical features are indicated by arrows and labeled with $\bar{E}_{0}$, ....$E_{1}'$ being assigned to peaks according to the band gaps calculated in the electronic band structure in Figure \ref{fig:snsoi-PBE-01-15-20-b} (a) without and with spin-orbit interaction. The energies of transitions in unstrained $\alpha-$Sn shown in Figure \ref{fig:asnldaso-6-22-18-w-exp} (a) and (b) are estimated using the method mentioned in \cite{gajdovs2006linear} and are presented in Table ~\ref{table:transition}, together with available experimental data. 
	
	The computed imaginary part $\epsilon_{\alpha \beta}(\omega)''$ of the frequency dependent dielectric function of $\alpha-$Sn, without and with  spin-orbit coupling calculated in the LDA scheme for exchange and correlation, are presented in Figure \ref{fig:asnldaso-6-22-18-w-exp} (a) and \ref{fig:asnldaso-6-22-18-w-exp} (b), respectively. In general, except for some deviations, the calculated imaginary part of dielectric functions $\epsilon_{\alpha \beta}(\omega)''$ agree well with the experimental measurement with respect to the line shape, peak positions, and peak intensities. \cite{carrasco2018direct}  The following discussion is concentrated on the spin-orbit effect.
	
	Figure \ref{fig:asnldaso-6-22-18-w-exp} (b) shows that spin-orbit interaction has effects on $\epsilon''$ of $\alpha-$Sn, in particular, in lower energy region, in three respects. First, a finer lineshape is obtained. The calculated dielectric function captures transitions such as $\bar{E}_{0}^{\prime}$, $\bar{E}_{0}^{\prime}+\Delta_{0}^{\prime}$, which are not clearly observed in the ellipsometric measurements shown in red line in Figure \ref{fig:asnldaso-6-22-18-w-exp} (b). Second,  calculated $E_{1}$ and $E_{1}+\Delta_{1}$ peaks agreement better with the ellipsometric measurements. 	We calculated $E_{1}$ and $E_{1}+\Delta_{1}$ to be 1.30 eV and 1.78 eV, close to our experimental value 1.278 eV and 1.73 eV, respectively.\cite{carrasco2018direct}  But, The calculated $\bar{E}_{0}$ value is higher than our experimental value by 0.35 eV. The calculated $E_{2}$ is 3.26 eV, lower than our experimental value by 0.19 eV.\cite{carrasco2018direct}	Third, The magnitude of calculated peaks are comparable with these experimental values, with the largest difference in the $\bar{E}_{0}$ peak, whose magnitude is three times higher than the value without spin-orbit interaction. These changes may be because, as we discussed earlier, split-off band $\Gamma_{7v}^{+}$ modifies the band structure at $\Gamma$ point, where $\bar{E}_{0}$, $E_{1}$, and $E_{1}+\Delta_{1}$ transition occur (see $\bar{E}_{0}$ calculated without and with spin-orbit interaction in Figure \ref{fig:snsoi-PBE-01-15-20-b} (a) and (b), respectively). Moreover, the spin-orbit interaction induces the topological states, thus, altering the optical oscillator strength significantly by 20 \% as compared to the normal system. \cite{diaz2018topologically} The oscillator strength of $\bar{E}_{0}$ can be further strengthened by the hybridization of $p$ and $s$ orbits in $\Gamma_{7c}^{-} $ and $ \Gamma_{8v}^{+}$, as shown in Figure \ref{fig:snsoi-PBE-01-15-20-b}(b).  Base on these facts, it can be concluded that\ $\bar{E}_{0}$ energy peak is indeed from the interband transition between $\Gamma_{8v}^{+}  \rightarrow  \Gamma_{7c}^{-}$. \cite{lu2019modulated} Nevertheless, there is no remarkable change in band structure at $X$ point, thus, $E_{2}$ peak is unchanged. The inclusion of spin-orbit interaction leads a better agreement with experimental measurement and the following discussion focuses on the results obtained from HSE functional. The dielectric function of $\alpha-$Sn is also calculated using HSE functional with spin-orbit interaction and presented in Figure \ref{fig:asn-hse-exp-190528}. The major difficulty of HSE functional is that the calculation is computationally demanding and slow to converge. To overcome the difficulty, the calculation is performed with improved dielectric method to moderate the number of $\mathbf{k}$ points to accelerate converging. In consequence, the calculated dielectric function demonstrates wiggles and higher magnitude specifically $E_{1}$ and $E_{1}+\Delta_{1}$. Their positions are red-shift compared with experimental results and this seems to agree with the trend shown in Figure \ref{fig:asnldaso-6-22-18-w-exp} (a).  Peaks $\bar{E}_{0}$ and $E_{2}$ demonstrate clearly. The estimated $\bar{E}_{0}$ is 0.480 eV, which is close to our experimental value of 0.416 eV. Our result agrees with the argument that HSE is more appropriate for calculating optical gaps, in particular, for $\bar{E}_{0}$ peak. \cite{brothers2008accurate} The significant difference in calculated $\bar{E}_{0}$ value (representing the transition  $\Gamma_{8v}^{+} \rightarrow \Gamma_{7c}^{-}$) with LDA and HSE hybrid functional can be understood from the different band order of $\Gamma_{8v}^{+}$, $\Gamma_{7v}^{+}$, and $\Gamma_{7c}^{-}$ near the $\Gamma$ point in the band structure, as shown in Figure \ref{fig:snsoi-PBE-01-15-20-b} and \ref{fig:snsoi-HSE-band-05-15-19}, where LDA overestimates the spin-orbit interaction as discussed before.  LDA can predict the dielectric function of $\alpha$-Sn with a good agreement with the experimental measurement, except overestimating the peak $\bar{E}_{0}$ by 0.36 eV. In contrast, HSE captures the $\bar{E}_{0}$ peak precisely, but underestimates the $E_{1}$ and $E_{1}+\delta_{1}$ by 0.15 and 0.24 eV, respectively. 
	
	\begin{table}
		\caption{Energies (in eV) of peaks in $\alpha-$Sn extracted from the imaginary part dielectric function shown in Figure \ref{fig:asnldaso-6-22-18-w-exp} and \ref{fig:asn-hse-exp-190528}. The values are compared with available experimental results.}	
		\label{table:transition} 		
		\begin{center}
			\begin{ruledtabular}
				\begin{tabular}{ c c c c c c  }	
					Energies               & \multicolumn{3}{c}{This work}    & Exp. \footnotemark[1] & Exp. \footnotemark[2] \\ \cline{2-4}		
					& LDA (w.SO)    & LDA     & HSE   &                       &                      \\                    
					\hline 
					$\bar{E}_{0}$              &  0.769     & 0.30    & 0.480 &  0.410                & $\mathrm{-}$               \\
					$E_{1}$                &  1.30      & 1.00     & 1.129  & 1.278                 & 1.365                \\      
					$E_{1}+\Delta_{1}$     &  1.78      & 1.48     & 1.489  & 1.73                  &  1.845               \\
					$E_{2}$                &  3.26      & 3.10      & 3.602      & 3.45                  & 3.718                \\
					%$E_{1}^{\prime}$ & 4.63 & - &  4.11 & $L_{6c}^{+}  \rightarrow  L_{4,5v}^{-}$  \\
					%$E_{1}^{\prime}+\Delta_{1}^{\prime}$ & 4.83 & - & 4.39 & $L_{4,5c}^{+}  \rightarrow  L_{4,5v}^{-}$  \\
					%$E_{1}^{\prime}+\Delta_{1}$ & 4.74 & - & - & $L_{6c}^{+}  \rightarrow  L_{6v}^{-}$  \\
					%$E_{1}^{\prime}+\Delta_{1}^{\prime}+\Delta_{1}$ &  5.00 & - & 4.89 & $L_{4,5c}^{+}  \rightarrow  L_{6v}^{-}$  \\
				\end{tabular}
			\end{ruledtabular}
			\footnotetext[1]{The experimental data from Ref.~\cite{carrasco2018direct}}
			\footnotetext[2]{The experimental data from Ref.~\cite{cardona1966electroreflectance}}
		\end{center}
	\end{table}

\begin{figure}[h]
		\centering
		\includegraphics[width=8cm, height=6cm]{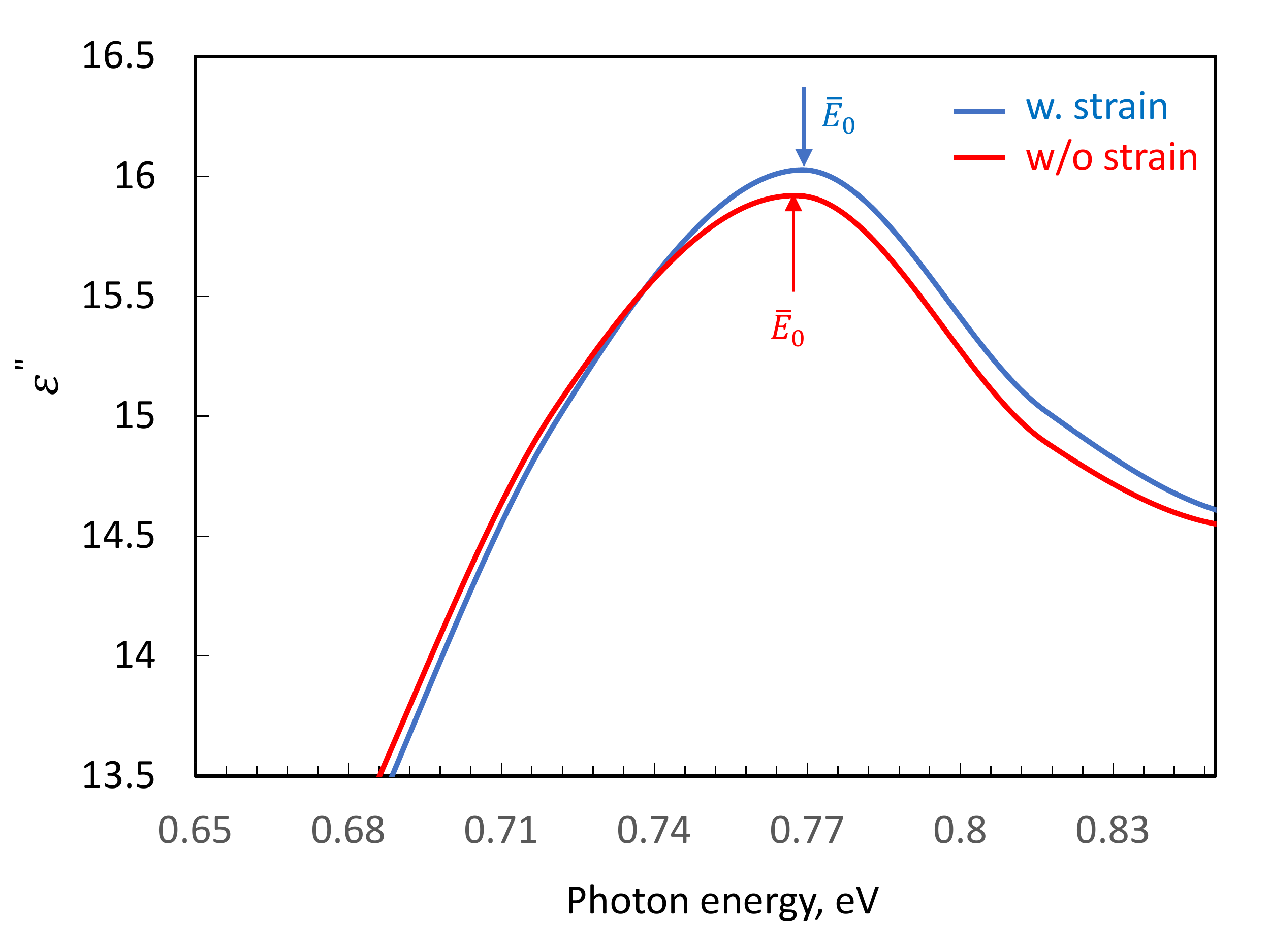}
		\caption{Comparison between the computed frequency dependent the imaginary part of the dielectric function of $\alpha$-Sn using LDA with and without an in-plane compressive strain of $-0.164\%$, shown in red and blue line, respectively. 
		}
		\label{fig:asnldasos-6-25-18-compare-b}
	\end{figure}

	The dielectric function of $\alpha-$Sn around $\bar{E}_{0}$ with $- 0.164$\% in-plane compressive strain is calculated and presented together with results without strain in Figure \ref{fig:asnldasos-6-25-18-compare-b}. The strain increases slightly the peak $\bar{E}_{0}$ by $\sim$ 1.5 meV, which agrees with our recent ellipsometric measurement that showed critical points by 5 meV by in-plain strain of $-0.13 \%$ using deformation formalism.\cite{carrasco2018direct} More importantly, it provides evidence that the $\bar{E}_{0}$ transition does not arise from externally applied strain, but instead from a direct interband transition $\Gamma_{8v}^{+} \rightarrow \Gamma_{7c}^{-}$. 
	
	\begin{figure}[h]
		\centering
		\includegraphics[width=8.3cm, height=6cm]{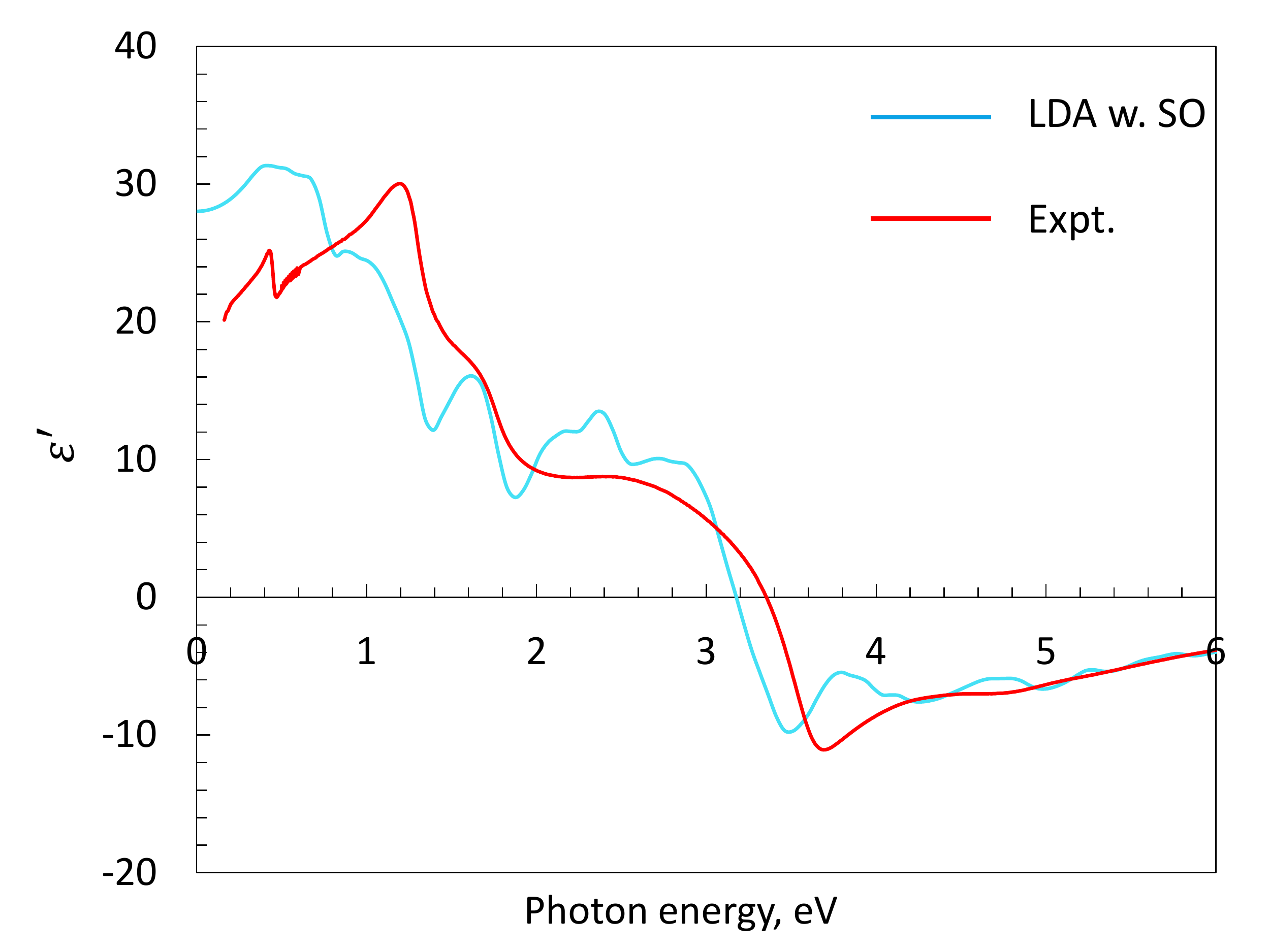}
		\caption{The computed the real part of the frequency dependent dielectric function of $\alpha$-Sn using LDA with the spin-orbit interaction shown in cyan compared with the experimental measurement shown in red. \cite{carrasco2018direct}} 	
		\label{fig:asnldaso-6-22-18-real}
		%			asnldaso-6-22-18-real.jpg
	\end{figure}
	
	The real part of the dielectric function $\epsilon^{\prime}$ is shown in Figure \ref{fig:asnldaso-6-22-18-real} along with the experimentally determined $\epsilon'$.  A sum rule \cite{peter2010fundamentals} is used to estimate $\epsilon_{\infty}$ as 23.4, \cite{carrasco2018direct} which is close to $\epsilon_{\infty}=24$ measured by infrared polarized reflectance. \cite{lindquist1964optical} The calculated dielectric constant $\epsilon_{\infty}$ is 28, which is larger than the experimental value by 4. The $\epsilon^{\prime}$ becomes negative at 3.2 eV and continue to the minimum point at 3.45 eV, which relates to $E_{2}$ transition.  The other optical properties of $\alpha$-Sn, such as refractive index and absorption spectrum, which are mainly naturally derived from the dielectric constants are presented in the appendix \ref{sec:appendix-1} and \ref{sec:appendix-2}, respectively for reference.

	\section{\label{sec:conclusion}Conclusion}
	In this paper, we explore the dielectric function of $\alpha-$Sn both unstained and strained with a focus on the infrared wavelength (photon energy $<$ 1 eV) using the first-principles method. The optical transition at 0.41 eV observed in the our elliposometric experiment originates from the interband transition $\Gamma_{8v}^{+} \rightarrow \Gamma_{7c}^{-}$ and is explained by combining first-principles calculation of the band structure and dielectric function. In the strained $\alpha-$Sn, this asymmetric optical excitation from the occupied to the unoccupied states across the Dirac point in the helical Dirac cone, producing spin-polarized surface current. This conclusion could provide a foundation for further study of the spin-polarized photocurrent in strained $\alpha-$Sn, a topological elemental material, and for its applications in high-speed communications and opto-electronic devices. Validation and justification of the exchange-correlation functionals are very important in calculation of the electronic and optical properties of the topological materials. HSE can capture the interband excitation in the helical Dirac cone with a reasonable accuracy, but LDA overestimates this transition.

	\appendix
	\section{\label{sec:appendix-0}Topological surface states in strained $\alpha$-Sn}
	
\begin{figure}[h]
		\centering
		\includegraphics[width=8.5cm, height=4.8cm]{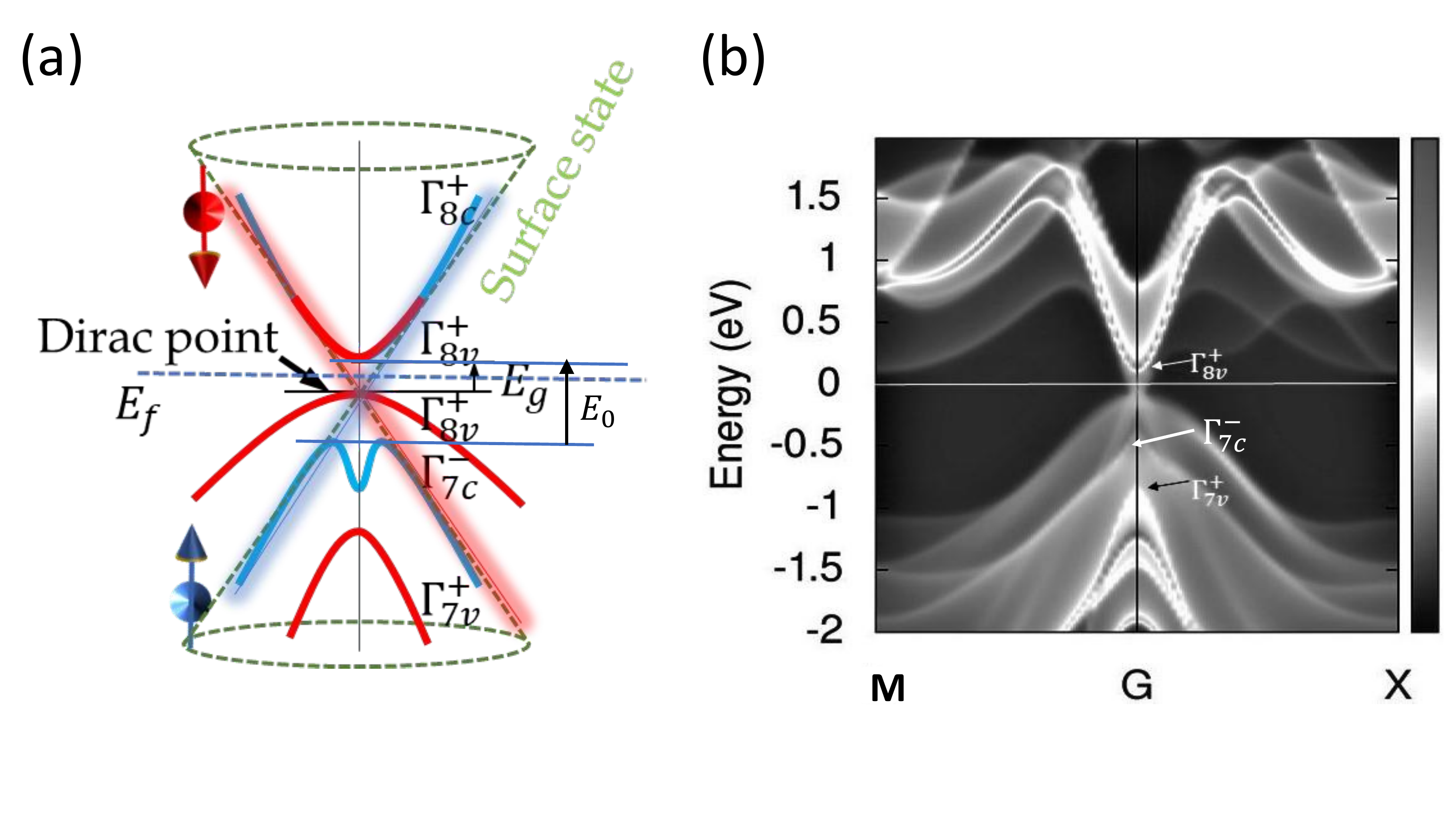}
		\caption{The surface states illustrated (a) and calculated by HSE (b) in strained $\alpha$-Sn, a  topological elemental material.
		}
		\label{fig:asn-surface-08-01-22}
	\end{figure}

	The helical surface states in $\alpha$-Sn are calculated with HSE and shown in Figure \ref{fig:asn-surface-08-01-22}. The optical transition  $\Gamma_{8v}^{+} \rightarrow \Gamma_{7c}^{+}$ across the Dirac point, which is below the Fermi level, in the helical Dirac cone.
	
	\section{\label{sec:appendix-1}The refractive index of $\alpha$-Sn}
	
\begin{figure}[h]
		\centering
		\includegraphics[width=8.2cm, height=6.35cm]{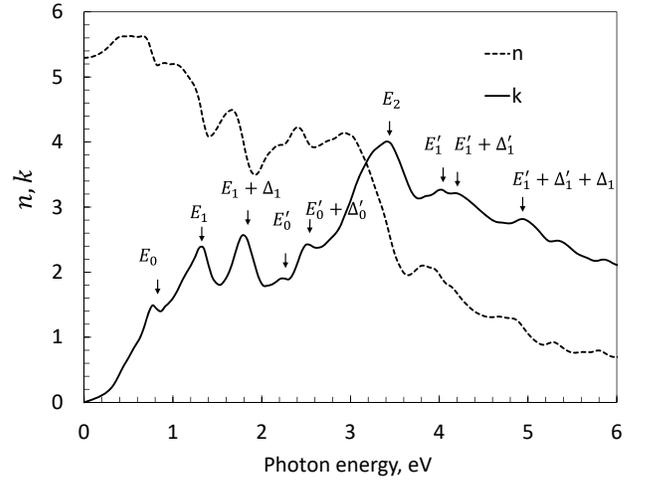}
		\caption{The computational refractive index $n$ (dashed line) and extinction coefficient $k$ (solid line) of $\alpha$-Sn.
			%aSn-dielectric-35-1-17-18-ref.jpg
		}
		\label{fig:asnldaso-6-22-18-nk}
	\end{figure}

	The refractive index ($n$) and extinction coefficient ($k$) is calculated for $\alpha$-Sn using the following equations \cite{adachi1989optical} and demonstrated in Figure \ref{fig:asnldaso-6-22-18-nk}.
	
	\begin{equation}\label{eq:refractive-1}
		n=\Bigg(\frac{(\epsilon^{\prime 2}+\epsilon^{\prime\prime 2})^{1/2}+\epsilon^{\prime}}{2}\Bigg)^{1/2},
	\end{equation}
	
	\begin{equation}\label{eq:refractive-2}
		k=\Bigg(\frac{(\epsilon^{\prime 2}+\epsilon^{\prime \prime 2})^{1/2}-\epsilon^{\prime}}{2}\Bigg)^{1/2},
	\end{equation}

	where $\epsilon^{\prime}$ and $\epsilon^{\prime \prime}$ are taken from the dielectric function computed using LDA with inclusion of spin orbit interaction, as discussed in section \ref{sec:dielectric}. The extinction coefficient $k$ demonstrates distinct peaks associated with transitions, which are labeled in Figure \ref{fig:asnldaso-6-22-18-nk}. For example, the peak at 0.71 eV corresponds to $\bar{E}_{0}$ transition. 
	
	\section{\label{sec:appendix-2}The absorption spectrum of $\alpha$-Sn}
	
	\begin{figure}[h]
		\centering
		\includegraphics[width=8.2cm, height=6.15cm]{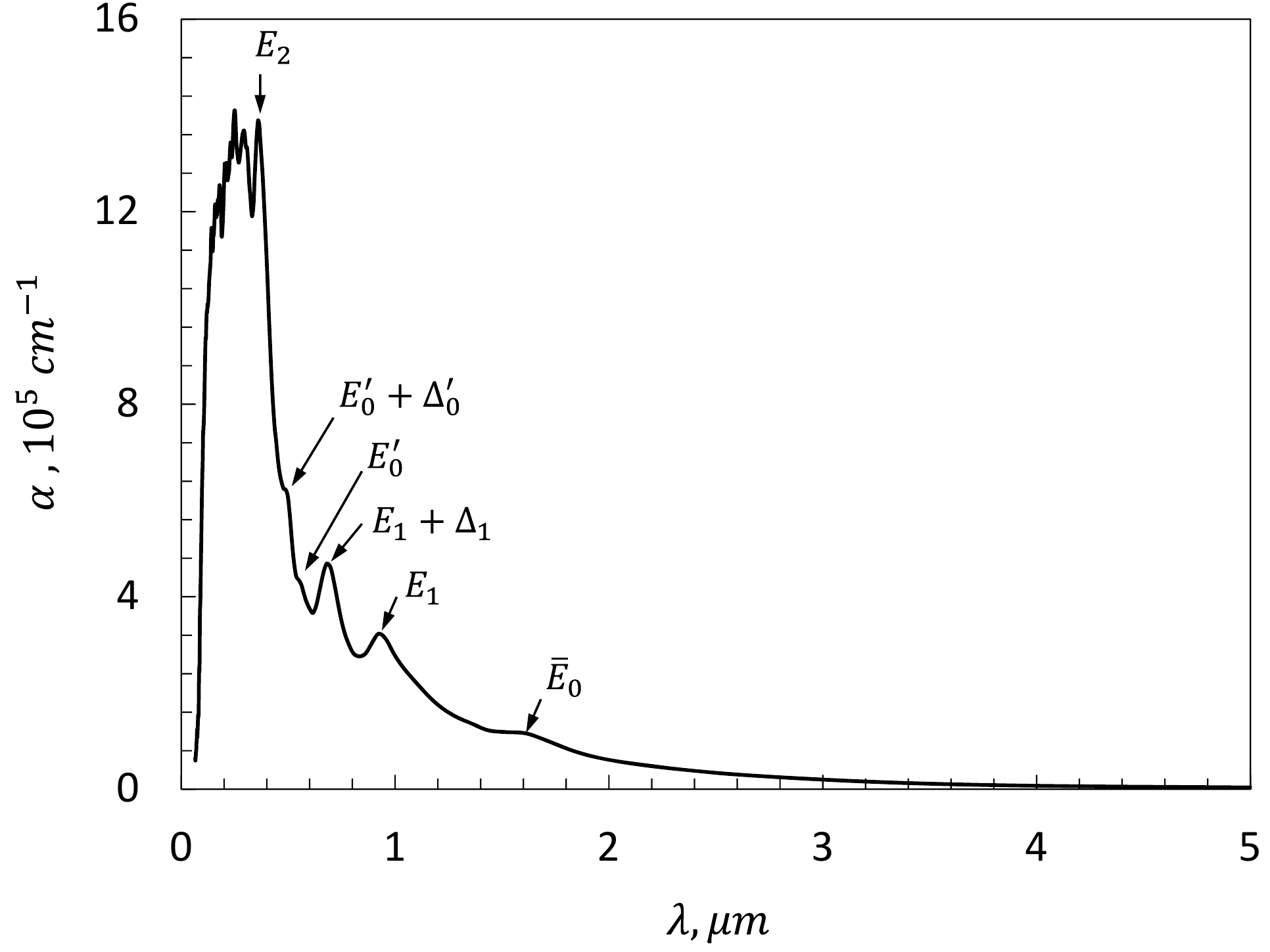}
		\caption{Computational absorption spectrum of $\alpha$-Sn as a function of wavelength. 
			%aSn-dielectric-35-1-17-18-asb-3.jpg
		}
		\label{fig:asnldaso-6-22-18-abs}
	\end{figure}
	
	The absorption coefficient $\alpha$ is calculated using a relation: \cite{adachi1989optical}
	\begin{equation}\label{eq:absorption-1}
		\alpha=4\pi k/\lambda,
	\end{equation}
	
	where $k$ is the extinction coefficient obtained with equation \ref{eq:refractive-2} in section \ref{sec:appendix-2}, $\lambda$ is the wavelength. The calculated absorption spectrum of $\alpha$-Sn is shown in Figure \ref{fig:asnldaso-6-22-18-abs}.

	The absorption spectrum captures the critical points in the long wavelength region. In particular, the absorption peak at 1.614 $\mu m$ corresponds to transition $\bar{E}_{0}$ ($\Gamma_{7}^{-}$ to $\Gamma_{8}^{+}$).  Transitions $E_{1}$, $E_{2}$, and also the transitions due to spin orbit spiting, $E_{1}+\Delta_{1}$, $^{\prime}$, $^{\prime}+\Delta_{0}^{\prime}$, are clearly obvious.
	
	The absorption spectrum of $\alpha$-Sn also provide the information of the band structure. $\alpha$-Sn becomes a direct band gap semiconductor in the long wavelength region because an absorption peak at 1.614 $\mu m$ is observed. The transition $E_{1}$ and $E_{1}+\Delta_{1}$ are indirect because both conduction and valence band edges are not at the center of the Brillouin zone. 
	
	\begin{acknowledgments}
		J. S. D. wishes to thank Air Force Office of Scientific Research for financial support and DoD High Performance Computing Modernized Program for providing computational support.  
	\end{acknowledgments}

	\nocite{*}
	
	\bibliography{apssamp}% Produces the bibliography via BibTeX.
	
\end{document}